\documentstyle[prd,aps]{revtex}
\begin{document}
\twocolumn[\hsize\textwidth\columnwidth\hsize\csname @twocolumnfalse\endcsname
\title{
     A solution to the problem posed by Byland and Scialom
}
\author{
     M. Yu.\ Zotov
}
\address{
     D. V. Skobeltsyn Institute of Nuclear Physics, \\
     Moscow State University, Moscow 119899, Russia
}
\date{e-mail: \tt zotov@eas.sinp.msu.ru}
\maketitle
\begin{abstract}
     Recently, Byland and Scialom studied the evolution of the Bianchi~I,
     the Bianchi~III and the Kantowski--Sachs universe on the basis of
     dynamical systems methods ({\it Phys.\ Rev.\ }{\bf D57}, 6065 (1998),
     gr-qc/9802043).  In particular, they have pointed out a problem to
     determine the stability properties of one of the degenerate critical
     points of the corresponding dynamical system.  Here we give a
     solution, showing that this point is unstable both to the past and
     to the future.  We also discuss the asymptotic behavior of the
     trajectories in the vicinity of another critical point.
\end{abstract}

]
\narrowtext
     In one of their recent works, Byland and Scialom studied the
     evolution of the Bianchi~I, the Bianchi~III and the
     Kantowski--Sachs universe in a model with a real scalar field and
     a convex positive potential~\cite{BS}.  A considerable part of the
     investigation was devoted to the analysis of the asymptotic behavior
     and the stability properties of the solutions of the
     Einstein--Klein--Gordon equations
\begin{eqnarray*}
     && \dot{\theta} = -\frac{1}{3} \, \theta^2
                       -2 \sigma^2 + V(\varphi) - \psi^2,             \\
     && \dot{\sigma} = -\frac{1}{3 \sqrt{3}} \, \theta^2
                       - \theta \sigma
                       + \frac{1}{\sqrt{3}} \left(\sigma^2
                         + V(\varphi) + \frac{1}{2} \, \psi^2 \right),
                                                                      \\
     && \dot{\varphi}= \psi ,                                        \\
     && \dot{\psi}   = -\theta \psi - \frac{dV}{d \varphi},
\end{eqnarray*}
     where $\theta$ is the function of the expansion rate,
     $\sigma_{\mu\nu}$~is the shear tensor of the hypersurface of
     constant time,
     $\sigma=\frac{1}{2} \, \sigma_{\mu\nu} \sigma^{\mu\nu}$,
     $\varphi$~is the scalar field, and~$V$ is
     a convex positive potential; an overdot stands for derivatives
     with respect to~$t$ (see Eqs.\ (8)--(11) in~\cite{BS}).

     The analysis was based on determining the stability properties
     of the critical points of the dynamical system
\begin{eqnarray} \label{MainDS}
     && S' = - \frac{1}{3 \sqrt{3}} - \frac{2}{3} S
            + \frac{1}{2 \sqrt{3}} (2 S^2 + 2 U^2 + P^2)
            + F S,                                          \nonumber \\
     && U' = \left(\frac{1}{3} - \frac{\lambda}{2} P + F \right) U,    \\
     && P' = -\frac{2}{3} P + \lambda U^2 + F P,             \nonumber
\end{eqnarray}
     where $S = \sigma/\theta$, $U = \sqrt{V}/\theta$,
     $P = \psi/\theta$, $F = 2 S^2 - U^2 + P^2$, $\lambda$ is given by
     $V = V_0 \, e^{-\lambda \phi}$, and a prime stands for derivatives
     with respect to~$\tau$ defined by $d\tau = \theta \, dt$ (see
     Eqs.\ (24)--(26) in~\cite{BS}).  Here we omit an equation
     for~$\theta$, since Eqs.\ (\ref{MainDS}) do not contain this
     function.

     It was found in~\cite{BS} that the dynamical
     system~(\ref{MainDS}) has the following critical points:
\begin{eqnarray*}
     &P_1:& \quad S = -\frac{1}{2 \sqrt{3}}, \; U = 0, \; P = 0  \\
     &P_2:& \quad S = 0, \; U = \sqrt{\frac{6 - \lambda^2}{18}}, \;
            P = \frac{\lambda}{3}, \quad \lambda \le \sqrt{6}    \\
     &P_3:& \quad S = \frac{1}{2 \sqrt{3}}
                     \frac{2 - \lambda^2}{1 + \lambda^2},        \;
            U = \frac{\sqrt{2 + \lambda^2}}{\sqrt{2}\,(1 + \lambda^2)},
            \;
            P = \frac{\lambda}{1 + \lambda^2}                    \\
     &\Sigma:& \quad U = 0, \; 3 S^2 + \frac{3}{2} P^2 = 1, \quad
            S \in [-1/\sqrt{3}, 1/\sqrt{3}].
\end{eqnarray*}
     In particular, it was shown that the point~$P_2$ has the
     eigenvalues $\varepsilon_1 = -1 + \lambda^2/6$ (twice) and
     $\varepsilon_2 = -2/3 + \lambda^2/3$.  Thus, it is stable
     for $\lambda < \sqrt{2}$ and unstable for
     $\sqrt{2} < \lambda < \sqrt{6}$.  It was also found that~$P_2$
     is also unstable for $\lambda = \sqrt{2}$.
     The problem is to determine the stability properties of~$P_2$
     for the case $\lambda = \sqrt{6}$, in which~$P_2$ is degenerate
     with the eigenvalues $\varepsilon_1 = 0$ (twice) and
     $\varepsilon_2 = 4/3$.  Here we shall demonstrate that the
     point~$P_2$ is unstable (both to the future and to the past).


     Notice that for the case $\lambda = \sqrt{6}$, $P_2$ belongs to
     the ellipsis~$\Sigma$.  Thus, one of the zero eigenvalues
     corresponds to the fact that $\Sigma$ is a one-dimensional critical
     set.  In order to prove that the critical point~$P_2$ is unstable
     both to the past and to the future, it is sufficient to show that
     there is a projection of this point, which is unstable.   To do
     this, we rewrite the dynamical system~(\ref{MainDS}) as
\begin{eqnarray} \label{ShDS4P2}
     && S' = 2 \alpha
              \left( \frac{1}{2 \sqrt{3}} + S \right) \bar{P}
             + \frac{1}{2 \sqrt{3}} (2 S^2 + 2 U^2 + \bar{P}^2)
             + \bar{F} S,                              \nonumber     \\
     && U' = \left( \frac{\alpha}{2} \, \bar{P} + \bar{F} \right) U, \\
     && \bar{P}' = \frac{4}{3} \bar{P}
             + \alpha \, (3 U^2 + 2 \bar{P}^2)
             + (\alpha + \bar{P}) \, \bar{F},          \nonumber
\end{eqnarray}
     where $\bar{P} = P - \alpha$, $\alpha = \sqrt{2/3}$, and
     $\bar{F} = 2 S^2 - U^2 + \bar{P}^2$.  The point~$P_2$ now
     corresponds to the origin, $(S,U,\bar{P}) = (0,0,0)$.

     Consider the projection of~(\ref{ShDS4P2}) in the plane $S = 0$.
     The equations for~$U$ and~$\bar{P}$ may now be written as
\begin{mathletters} \label{eqs4S=0}
\begin{eqnarray}
     && U' =  \left( \frac{\alpha}{2} \, \bar{P}
              - U^2 + \bar{P}^2 \right) U,              \label{e4U}  \\
     && \bar{P}' = \frac{4}{3} \bar{P}
                  + (2 \alpha - \bar{P}) \, U^2
                  + (3 \alpha + \bar{P}) \, \bar{P}^2.  \label{e4P}
\end{eqnarray}
\end{mathletters}
     Denote the right hand sides of Eqs.~(\ref{e4U}) and~(\ref{e4P})
     as ${\cal U}(U,\bar{P})$ and ${\cal P}(U,\bar{P})$ respectively.
     The idea is to find a solution $\bar{P} = f(U)$ of the
     equation ${\cal P}(U,\bar{P}) = 0$ in a neighborhood of $U=0$,
     to substitute it into~${\cal U}(U,\bar{P})$:
\[
     {\cal U}(U,f(U)) = a_m U^m + \dots ,
\]
     and then to examine whether the power~$m$ of the leading term is
     even or odd and, in the latter case, to check the sign of~$a_m$
     (see, e.g.,~\cite{Perko}).

     It is easy to see that the corresponding solution in our case
     assumes the form
\[
     \bar{P} = f(U) = - \sqrt{\frac{3}{2}} \, U^2 + O(U^4).
\]
     Therefore,
\[
     {\cal U}(U,f(U)) = - \frac{3}{2} \, U^3 + O(U^7).
\]
     Thus, $m = 3$ is odd, and $a_m$ is negative.  It follows immediately
     from the standard results for the two-dimensional dynamical systems
     that the point $(U,\bar{P}) = (0,0)$ is a topological saddle. Hence,
     it is unstable both to the future and to the past.


     Let us mention that the same conclusion may be obtained basing on
     the results of the center manifold theory (see, e.g.,~\cite{Wiggins}).


     We remark that~$P_2$ is not the only degenerate point of~$\Sigma$.
     Recall that the eigenvalues of~$\Sigma$ are~\cite{BS}
\[
     \varepsilon_1 = 0,                                          \quad
     \varepsilon_2 = 1 - \frac{\lambda}{2} \, P,                 \quad
     \varepsilon_3 = \frac{2}{3} (2 + \sqrt{3} \, S).
\]
     Thus, for any point of~$\Sigma$, except for the points
     $(P,S) = (0, \pm 1/\sqrt{3})$, there exists $\lambda = 2/P$,
     $\lambda \in [-\sqrt{6}, \sqrt{6}]$ such that~$\varepsilon_2$
     turns to zero, thus making this point degenerate.  The stability
     properties of these degenerate points may be studied in a way
     similar to the above.


     Finally, let us make some comments on the behavior of the
     trajectories of~(\ref{MainDS}) in the vicinity of the point~$P_1$.
     It was found in~\cite{BS} that this point has the eigenvalues
     $\varepsilon_{S,P} = -1/2$ and $\varepsilon_U = 1/2$.  It was also
     claimed that starting around~$P_1$, the critical point can never
     be reached by any solutions of the dynamical system.  This is not
     quite correct.  It follows from the standard results of the
     dynamical systems theory (see, e.g.,~\cite{Hartman}, Theorem~6.1)
     that in a neighborhood of~$P_1$ there exists a two-dimensional
     stable manifold~$W^s$ and a one-dimensional unstable
     manifold~$W^u$.  The trajectories lying on these manifolds tend
     to~$P_1$ as $\tau \to \infty$ and $\tau \to -\infty$
     respectively.  One can easily find the asymptotic behavior of
     these solutions.

     Namely, let us introduce $\bar{S} = S + 1/(2 \sqrt{3})$.  Then
     the dynamical system~(\ref{MainDS}) reads as
\begin{eqnarray} \label{ShDS}
     && \bar{S}' = - \frac{1}{2} \bar{S}
            - \frac{1}{2 \sqrt{3}} \left(4 \bar{S}^2 - 3 U^2 \right)
            + \bar{F} \bar{S},                               \nonumber \\
     && U' = \frac{1}{2}
              \left(1 - \lambda P - \frac{4}{\sqrt{3}} \, \bar{S}
               + 2 \bar{F} \right) U,                                  \\
     && P' = -\frac{1}{2} P - \frac{2}{\sqrt{3}} \, \bar{S} P
            + \lambda U^2 + \bar{F} P,                       \nonumber
\end{eqnarray}
     where $\bar{F} = 2 \bar{S}^2 - U^2 + P^2$.  The point~$P_1$
     now corresponds to the origin, $(\bar{S},U,P) = (0,0,0)$.
     The eigenvectors of this point are $\zeta_{\bar{S}} = (1,0,0)$,
     $\zeta_P = (0,0,1)$, and $\zeta_U = (0,1,0)$.  Now one can see
     that the trajectories on~$W^s$ take the form
\[
     U \equiv  0, \quad
     S \approx -\frac{1}{2 \sqrt{3}} + C_S \, e^{-\tau/2},   \quad
     P \approx C_P \, e^{-\tau/2},
\]
     as $\tau \to \infty$, where $C_S$ and $C_P$ are arbitrary constants,
     $C_S^2 + C_P^2 \ne 0$.
     (One can also obtain these solutions in a parametric and partially
     in an explicit form (for $\bar{S} \equiv 0$).)
     Notice that~$S$ here is just a linear function of~$P$.
     We also mention that one of the trajectories on~$W^s$, namely,
\[
     U \equiv 0,                                                 \quad
     S = -\frac{1}{2 \sqrt{3}} + \frac{1}{2 \sqrt{2}} \, P       \quad
     \text{for } P \in \; ]0, \sqrt{2/3}[,
\]
     joins~$P_1$ to $P_2$ if $\lambda = \sqrt{6}$.

     In order to obtain the asymptotic behavior of the trajectories
     on~$W^u$, we notice that for these trajectories $\bar{S} = o(U)$ and
     $P = o(U)$ as $U \to 0$. This allows us to consider only the leading
     terms in~(\ref{ShDS}):
\[
     \bar{S}' = - \frac{1}{2}\, \bar{S} + \frac{3}{2}\, U^2, \quad
     U' = \frac{1}{2} \, U,                                  \quad
     P' = -\frac{1}{2}\, P + \lambda U^2.
\]
     It follows immediately that the outgoing trajectories take the form
\[
     U \approx C_U \, e^{\tau/2},                                   \quad
     S \approx -\frac{1}{2 \sqrt{3}} + \frac{1}{\sqrt{3}} \, U^2,   \quad
     P \approx \frac{2}{3} \, \lambda U^2,
\]
     as $\tau \to -\infty$, where $C_U$ is a nonzero constant.

\acknowledgments

     I would like to thank Professor D.~V.~Gal'tsov for attracting
     my attention to the theory of dynamical systems.


\end{document}